\documentclass[smallextended]{svjour3}
\pdfoutput=1
\usepackage{graphicx, bm, color}
\usepackage{hyperref}
\usepackage{amsmath}

\newcommand{\rd}[1]{{\color[rgb]{0,0,0.7}#1}}

\begin{document}
\title{Note on the Kaiser-Peacock paper regarding gravitational lensing effects}
\author{George F R Ellis  \and Ruth Durrer}
\institute{G F R Ellis \at
	Mathematics Department, University of Cape Town\\
	Rondebosch, Cape Town 7700, South Africa \\
	Tel.: +27-21-650-2339\\
	\email{george.ellis@uct.ac.za}
\and
R Durrer \at
	Universit\'e de Gen\`eve, D\'epartement de Physique Th\'eorique\\
	Quai E. Ansermet 28, 1211 Gen\`eve, Switzerland \\
	\email{ruth.durrer@unige.ch}}
\date{Received: date / Accepted: date}
\maketitle

\begin{abstract}
	\textit{This paper revisits the controversy concerning whether gravitational lensing effects make a significant difference to estimation of distance to the Cosmic Microwave Background last scattering surface in cosmology. A recent paper by Kaiser and Peacock \cite{Kaiser_Peacock} supports a previous paper by Weinberg \cite{Wei76} stating that such affects average to zero because of energy conservation. In this note, problems are pointed out in the Kaiser and Peacock analysis related 
		 to their choice of endpoint of integration, and  to the `wrinkly surface' argument.
	 } 
\end{abstract}

\section{Introduction}
Various papers such as \cite{Bertotti} claim that lensing due to inhomogeneity can significantly  affect number counts in cosmology. Weinberg \cite{Wei76} however argued that due to energy conservation, this is not the case: such effects will vanish when  one averages over the sky. In contrast  \cite{CMBdist} calculates  non-zero lensing effects relating to the effective distance to the Cosmic Microwave Background (`CMB') last scattering surface to second order in perturbation theory, which in effect contradicts Weinberg's result. In the paper  ``On the bias of the distance-redshift relation from gravitational lensing''  \cite{Kaiser_Peacock}, hereafter KP,  Kaiser and Peacock defend Weinberg's thesis in the face of the various papers challenging it, and give  detailed references to previous work on the issue. 
However problems arise with KP as regards the following issues:
\begin{itemize}
	\item Curve parameters
	\item Curve end-points
	\item The `wrinkly surface' argument.
\end{itemize}
Here we  discuss them in turn, after noting some basic formulae in the following section. The method proposed here (cf \cite{BE}) is as far as possible to work in the real inhomogeneous universe rather than the fictional background model, approximating the resulting equations as needed. Perturbations are defined in terms of the difference at each point in the inhomogeneous model between the real value of a quantity there, and the mapped value from the background model. 

\section{Preliminaries: CMB Temperature}
A key issue is where the CMB decouples from matter, thus determining the observed CMB temperature and redshift.

\subsection{Redshift} 
For a family of observers  with future directed 4-velocity $u^a$ ($u^b u_b =-1$)\footnote{The signature used here is $(-1,1,1,1)$.} one finds the basic redshift formula (\cite{Ehlers}; \cite{ell71}:section 6.2)
\begin{equation}\label{eq:redshift}
1+z = \frac{\lambda_\text{observed}}{\lambda_\text{emitted}} = \frac{(u_a k^a)_\text{emitter}}{(u_b  k^b)_\text{observer}} \,.
\end{equation}
We consider the past-directed null geodesic curve $x^a(v)$, where $v$ is an affine parameter, with tangent vector  $k^a$ pointing from the observer to the source:
\begin{equation}\label{eq:null_geodesic}
k^a = \frac{dx^a}{d v}: \,\,k^a k_a = 0, \,\,\, k^a_{\,\,\,; b}k^b = 0.
\end{equation}
If the parameter $v$ is not affine, one will get instead
\begin{equation}\label{eq:null_geodesic_non}
k^a k_a = 0, \,\,\, k^a_{\,\,\,; b}k^b = f(v) k^a
\end{equation}
for some function $f(v) \neq {0}$. We note here that if $k_a = \phi_{;a}$ for a scalar field $\phi$, it is automatically affine (\cite{ell71}:(6.4),(6.9)). Thus this will indeed be true for the tangent vector field to the past null cone $w = const$, because then $k_a = w_{,a}$ \cite{ell71}. But it will not automatically be true if one calculates along individual null geodesics or bundles of null geodesics with some geometrically chosen curve parameter $v$ such as cosmic time $t$ or distance travelled $d$. 

The direction $n^a$ of a past directed light ray relative to a (future directed) 4-velocity $u^a$ is  
\begin{equation}\label{eq:null direction}
k^a = (-u^a + n^a) (u^b k_b),\,\, n^a n_a = 1,\,\,  n^a u_a = 0. 
\end{equation}
and 
\begin{equation}\label{eq:dist}
dl = (u^a k_a) dv
\end{equation} 
is the projected spatial distance along the direction $n^a$ of the null vector corresponding to a curve parameter increment $dv$.

 The change of observed wavelength $\lambda$ down a past null geodesic for more and more distant sources is given by the derivative of (\ref{eq:redshift}) down the past null cone relative to an affine parameter and using the formula \cite{Ehlers} \cite{ell71}
\begin{equation}\label{eq:fluid}
u_{a;b} = \theta_{ab} + \omega_{ab} - \dot{u}_a u_b
\end{equation}
to give
\begin{equation}\label{eq:z_fluid}
\frac{d \lambda}{\lambda} = \{\theta_{ab} n^a n^b + (\dot{u}_a n^a)\} dl
\end{equation}
where $\theta_{ab}$ is the fluid expansion tensor, combining an isotropic expansion term $\theta$ and a shear term $\sigma_{ab}$, and $\dot{u}_a$ is the fluid acceleration \cite{Ehlers} \cite{ell71}. The first term is the Doppler shift term, with isotropic  and anisotropic parts, and the second the gravitational redshift term  (zero for a geodesic congruence). If the curve parameter is not affine,  relation (\ref{eq:z_fluid}) will have an extra term (the affine condition is assumed in its derivation, see Section 6.2.1 in \cite{ell71}).
 
\subsection{Temperature}
The surface intensity $I_\nu$ of radiation received from a source with spectrum ${\cal I}(\nu)$ and surface brightness $I_{G}(n^a)$ in the direction $n^a$ is given by  (\cite{ell71}:(6.40))
\begin{equation}\label{eq:SI}
I_\nu(n^a)d\nu  = I_{G}(n^a)\frac{{\cal I}(\nu)(1+z)d\nu}{(1+z)^3} 
\end{equation}
where the redshift is given by (\ref{eq:redshift}). This relation  
follows from reciprocity theorem (\cite{ell71}:section 6.4.3, \cite{Etherignton}), which is generically true. We can parametrise the direction $n^a$ as usual by spherical angles $(\theta,\phi)$. Applying (\ref{eq:SI}) to black body radiation,  it follows (\cite{ell71}:(6.41), (6.42)) that in each direction ($\theta,\phi$), radiation emitted as black body radiation at temperature $T_\text{emit}(\theta,\phi)$ at the point on the last scattering surface (LSS) in that direction is received as black body radiation at temperature $T_\text{obs}(\theta,\phi)$ where
\begin{equation}\label{eq:cmb}
T_\text{obs}(\theta,\phi) = \frac{T_\text{emit}(\theta,\phi)}{(1+z)}.
\end{equation}
Here $(\theta,\phi)$ denotes the direction of $n^\mu$ at the observer which equals the one at the emitter e.g. in geodesic light cone coordinates~\cite{Fanizza:2013doa}. 
This remarkable relation, resulting from a combination of results from quantum theory, statistical mechanics, and general relativity, is an exact  relation true in any cosmological spacetime.

\subsection{Decoupling}\label{decouple}
Decoupling of the CMB from matter takes place when the baryon temperature $T_m$ drops below the ionisation temperature $T_\text{dec}$ so that  negatively charged free electrons in the primordial plasma combine with positively charged nuclei to become neutral hydrogen and Thomson scattering ceases. Because the baryons and radiation are in equilibrium to a good approximation at that time, their temperatures are equal: 
\begin{equation}\label{eq:mattar_radn}
T_\gamma = T_m. 
\end{equation}
Decoupling therefore happens when the radiation temperature drops below $T_\text{dec}$; that is\footnote{Note that this is roughly the temperature at which  the density of photons with energy above the ionisation energy of hydrogen drops below the hydrogen density. The ionisation energy of hydrogen, $\sim 13.6$eV,  is much higher than $k_BT_\text{dec} \simeq 0.3$eV due to the high number of photons in the Universe, see~\cite{Durrer} for details.}, when
\begin{equation}\label{eq:decouple}
T_\gamma = T_\text{dec}.
\end{equation} 
Hence setting $T_\text{emit}(\theta,\phi) = T_\text{dec}$ in (\ref{eq:cmb}) the observed CMB temperature is  
\begin{equation}\label{eq:cmb1}
T_\text{obs}(\theta,\phi) = \frac{T_\text{dec}}{(1+z)}.
\end{equation} 
Now $T_\text{dec}$ is fixed by the physics of recombination of hydrogen (
we ignore the issue of the ionisation of helium) and $T_\text{obs}(\theta,\phi)$ is the measured CMB temperature, which varies over the sky. So what the measured CMB temperature in any direction tells us is the redshift $z_*(\theta,\phi)$ of the emission surface (the `cosmic photosphere') in that direction:
\begin{equation}\label{eq:cmb2}
(1+z_*(\theta,\phi))= \left({\frac{T_\text{dec}}{T_\text{obs}(\theta,\phi)}}\right)
\end{equation} 
where the numerator is given by the physics of ionisation and the denominator is the observed CMB temperature. 
As remarked by Sachs and Wolfe \cite{Sachs and Wolfe}, the interpretation of this redshift as being due to Doppler or gravitational  effects (e.g. whether it is due to the Rees-Sciama effect or local redshifts relative to the cosmological expansion)  does not affect this formula, which is completely general. The conclusion is
\begin{quote}
	\textbf{Lemma 1} \textit{Decoupling of matter and radiation (`the cosmic photosphere') takes place at the redshift $z=z_*(\theta,\phi)$ determined in terms of $T_\text{obs}(\theta,\phi)$ by (\ref{eq:cmb2}), irrespective of the causes of that redshift. Hence the correct boundary condition to use in evaluating lensing effects on the CMB is to determine the cosmic photosphere by setting $z = z_*(\theta,\phi)$ on the null geodesic in direction ($\theta,\phi$) for all $\theta,\phi$.}   
\end{quote}
The COBE, WMAP, and Planck images of the CMB are therefore just images of the variation of $z_*$ across the sky. \\
Here we have neglected the fact that decoupling it not instantaneous but  happens gradually, the finite depth of the decoupling surface and second order effects in the physics of decoupling such as the matter and radiation temperatures not being exactly equal,  and Helium decoupling effects. Even though the CMB anisotropy spectrum cannot be calculate with good precision when neglecting these effects, the main argument presented here remains valid.

That analysis depends on temperatures rather than densities. How does this relate to surfaces of constant density? For radiation, \begin{equation}\label{density_pert1}
\rho_\gamma = a T_{\gamma}^4\,\,\Rightarrow\,\, \delta\rho_\gamma = 4a T_{\gamma}^3 \delta T_{\gamma}
\end{equation}
and 
  pure adiabatic perturbations are characterised by
\begin{equation}\label{density_pert2}
\frac{\delta\rho_{\gamma}}{\rho_{\gamma}} = \frac{4}{3}\frac{\delta\rho_{m}}{\rho_{m}}\,\,\Leftrightarrow\,\, \delta\rho_{m} = 3\rho_{m} \frac{\delta T_{\gamma}}{T_\gamma}.
\end{equation}
Consequently in the adiabatic case,
\begin{equation}\label{density_pert3}
\{\delta T_\gamma = 0\} \Rightarrow \{\delta \rho_m = 0\}.
\end{equation}

\begin{quote}
	\textbf{Corollary 1} \textit{In the adiabatic case, by (\ref{eq:decouple}) and (\ref{density_pert3}) the LSS is a surface of constant baryon density $\delta \rho_m = 0$, so in that case the observed CMB fluctuations do not represent density fluctuations, as is often stated.}
\end{quote} 
  In standard perturbation theory language, this shows that in uniform density gauge (which for adiabatic perturbation is the same as uniform temperature gauge) the density fluctuations are given exactly by the redshift fluctuations. In the non-adiabatic case this will no longer be true. \\
  
As mentioned above,  the main shortcoming of the above analysis is of course the instantaneous recombination approximation (accurate to a few percent only for multipoles with $\ell<100$); to go beyond this one has to use a Boltzmann approach~\cite{Durrer}, but conceptually nothing changes.

\section{Curve parameter}
As pointed out above, the standard formula (\ref{eq:z_fluid}) assumes that the null geodesic is affinely parametrised.  A key issue then in doing CMB calculations is whether an affine parameter is used along the relevant geodesics, or some other parameter. Equivalently, what distance measure is chosen in the calculations? If it is for example chosen as comoving distance, however that is defined, relation (\ref{eq:z_fluid}) may no longer hold.

Eqn (1) in KP  is  the null geodesic focussing equation, which uses an  affine parameter, as does the geodesic deviation equation. However the caption to figure 1 says, ``\textit{In a hypothetical universe with inhomogeneity in some finite region of space, consider the mean fractional change to the area of a surface of constant redshift, or cosmic time}'', which  they compare with a surface of constant distance travelled. Now  a surface of constant time in a perturbed FLRW model is a gauge dependent quantity (it depends on the time parameter chosen in the inhomogeneous model), and there is no reason why it should be a surface of constant redshift, so these are generically two different surfaces. KP also refers to ``the radius reached by the light rays'' and a ``a path length $\lambda$'',  which are presumably both the integral of (\ref{eq:dist}) down the null geodesic. A variety of different distance measures are being used (they may be the same in the background model, but will not be so in the perturbed model).\\

 However KP then state (bottom of page below Figure 1)
\begin{quote}
	\textbf{KP Claim 1} \textit{: The rest of the paper consists of a calculation of the
		perturbation to the area of a surface of constant redshift.}
\end{quote}
As long as this is what is actually done, and this is confirmed at the start of Section 4.1 in KP, the confusion about what curve parameter is used need not matter: the cosmic photosphere is being treated as a surface of constant redshift. The perturbation to the area of that surface, with consequent changes in apparent distance, is due to gravitational lensing effects (the focussing equation and perhaps time delay effects) 
 as discussed in depth by KP. However if either the null focussing equation (equation (1) in KP) or the geodesic deviation equation is used to deduce area changes, then it does matter if the curve parameter is affine or not; and similarly if (\ref{eq:z_fluid}) or an equivalent equation is used to deduce the change of redshift down a family of null geodesics, then it matters as well.  

\section{ Curve end-point}
 Following Weinberg, the curve endpoint in KP is taken (see KP Claim 1)  as being on a surface of constant redshift, as in Figure 1 of \cite{Kaiser_Peacock}. But Section \ref{decouple} above implies this cannot be correct when examining CMB lensing,  if we use the correct physical conditions for decoupling. 
 
 \begin{quote}
 \textbf{Corrollary 2}\textit{ Equation	(\ref{eq:cmb1}) shows that if one calculates the CMB temperature $T_\text{calc}(\theta,\phi)$ using as endpoint a surface of constant redshift $z_\text{const}$ with correct physical conditions for decoupling, then $T_\text{calc}(\theta,\phi)$ will have no angular variation whatever: }
  \begin{equation}\label{eq:cmb11}
 T_\text{calc}(\theta,\phi) = \frac{T_\text{dec}}{(1+z_\text{const})}
 \end{equation}
 \textit{where the RHS is constant.
 }
 \end{quote}
 Hence taken at face value, the KP calculation does not give what is needed for CMB calculations in a perturbed FLRW model, where the observed CMB temperature varies due to varying redshifts of emission in different directions of observation in the sky. 
 \begin{quote}
 	\textbf{Corollorary 3} \textit{The observed CMB anisotropy $T_\text{obs}(\theta,\phi)$ is due to the difference in redshift in the inhomogeneous universe between a chosen reference such as a surface of constant redshift $z=z_\text{const}$ and the physical surface of decoupling $z=z_*(\theta,\phi)$ given by  (\ref{eq:cmb2})}.
 \end{quote} 
This difference for example determines all the anisotropies detected by  the Planck satellite observations. The issue is in fact acknowledged in Appendix A2 of KP, but they do not explain how they resolve it.\footnote{They say in A2, "\textit{The surface of last scattering	is in reality a surface of constant temperature but varying	redshift. Nevertheless, we can ask what temperature  fluctuations would be observed if we were able to see a surface of constant redshift, and the answer is that the observed CMB would be the same}''. The text does not however in transparent fashion explain how they determine  from the physics of decoupling the expected  CMB fluctuations on the surface of constant redshift.} But that is the heart of the physical effect.  It is calculated in detail by Durrer \cite{Durrer} within linear perturbation theory on a  background cosmology, and in \cite{CMBdist,Clarkson1} at second order in perturbation theory. This approach is satisfactory in the linear case, but becomes very opaque in the non-linear case when one mixes a variety of distance measures as KP do (in A2: redshift, in A3 and A4: distance along the light ray, in A5: optical path length and  conformal path length, finally in A6: redshift, as per KP Claim 1). This approach contrasts with the view proposed here where, as in \cite{BE}, one works as far as possible in the real inhomogeneous universe, and uses the physics of decoupling to determine the integration endpoint.  
  
\section{The `wrinkly surface' argument}\label{sec:wrinkle}

KP state that time delays cause a further effect: ``\textit{the surface is `wrinkled' owing to time
	delays induced by the density 
fluctuations ...
one can draw an analogy with the surface of a swimming pool
perturbed by random waves of small amplitude. These cause a
fractional increase in the area of the surface that is on the order
of the mean square tilt of the surface}'', which they call the `wrinkly surface' argument.\\

However one must take into account Sach's shadow theorem \cite{Sachs:1961}, \cite{Pirani} (see note a) in Section 6.4.1 of \cite{ell71}), which states that the shape and area of an image in a screen orthogonal to the light ray are independent of the velocity of an observer. To show this, consider a vector $x^a$ lying in a screen orthogonal to $k^a$, which is effectively what the LSS is for the observer; then \begin{equation}\label{eq:orthog}
x^ak_a = 0.
\end{equation} The screen is set perpendicular to the incoming light ray, else there will be projection effects simply due to the screen being at an angle relative to the direction of observation, as opposed to any effects caused by time delays, which are equivalent to different distances down the light cone. Then changing distance down the null cone by a parameter distance $dv$ (that is, a time delay effect) adds a null increment 
\begin{equation}\label{eq:screen}
dx^a = k^a dv, \,\,k_a k^a = 0,\,\, 
\end{equation} to each such vector $x^a$, equivalent to a sum of a time displacement and a spatial displacement. 
  The increment (\ref{eq:screen}) does not alter condition (\ref{eq:orthog}):
 \begin{equation}\label{key}
 \{(x')^a = x^a + dv \, k^a,\,\, x^a k_a = 0\}\,\, \Rightarrow\,\, \{(x')^a k_a = 0\}. 
 \end{equation}
 Sachs \cite{Sachs:1961} pointed out that such an increment does not change any magnitude or shape in the image: by (\ref{eq:screen}) applied to $x^a$ and a similar screen vector $y^a$, 
\begin{equation}\label{key}
\{(x')^a = x^a + dv \, k^a,\,\, (y')^a = y^a + dv\, k^a\}\,\, \Rightarrow\,\, \{(x')^a (y')_a = x^a y_a \}. 
\end{equation}
In particular, $(x')^a (x')_a = x^a x_a$.
 This applies to the CMB case if we regard the LSS as the screen space relative to our observations.
\begin{quote}
\textbf{Lemma 2} \textit{Lengths, angles, and areas in a screen space are unchanged by altering that space by a small amount down the past light cone, adding an extra spatial displacement and time displacement that together represent a null displacement for each point in the screen space. 
}
\end{quote}
 On the view taken here, one compares the real decoupling surface in the inhomogeneous spacetime with the image in that spacetime of the decoupling surface in the  background model. Then the shadow theorem applies down the real past light cone in the inhomogeneous spacetime, and there is no such effect.  
\begin{quote}
	\textbf{Corollorary \rd{4}} \textit{Applying Lemma 2 to the  LSS  in the real physical spacetime as compared to the image in that spacetime of the LSS in the background model, there is no wrinkly surface effect for the CMB emission surface.}
\end{quote}
 KP by contrast  consider the issue in the background spacetime and find a non-zero result; but the shadow theorem also applies there, in regard  to the background lightcone. However the past light cone in the background spacetime does not correspond to the image of the physical light cone in the inhomogeneous  spacetime, as shown in Figure A1 in KP: the imaged light cone is not a null surface. There can consequently be such an effect resulting from the mapping between the physical spacetime and the background spacetime, which is of course a gauge dependent relation. Working this out one must again show how the relation between the real surface of decoupling and a fictitious one works, which requires identifying physically the real surface of decoupling, which is not a surface of constant redshift (in contrast to KP  Claim 1). It also requires taking fully into account the Minkowski geometry that leads to the Shadow Theorem; it is not clear that KP does this. 
 
  The view of this note is that it would be clearer to consider such effects in the physical spacetime, where it vanishes.

\section{Conclusion}
The issues discussed in the previous sections raise queries about  KP. Further study needs to be done to see what changes this might cause to their conclusions as regards lensing effects and the CMB: how large might any such effect be? Other papers carefully considering second order perturbations \cite{CMBdist,Clarkson1,Clarkson2} reach different conclusions than they do. For the moment, we simply note the queries raised here imply the results of those papers are not obviously negated by KP. It is possible the results in KP are correct, but they have not been presented in a way that makes this clear.\\
 
 How does this relate to other CMB lensing studies, such as the major  study  by Lewis and Challinor \cite{Lewis Challinor}? Using linearised theory in the background spacetime, they emphasize just after their equation (2.10) that one can regard that equation either as giving a radial displacement or a time delay. However it is of course both (one is integrating down the past light cone), which is why Sach's Shadow Theorem applies.  They then integrate to a surface of constant time in the background model (a conformal time $\eta_0 - \eta_* = \chi_*$), whose image in the perturbed model will not be the same as the real surface of decoupling in the inhomogeneous spacetime as given by (\ref{eq:cmb2}). This difference does not matter in the linear case. The issue at hand is whether the difference between these two integration limits causes a detectable effect in precision measurement of the CMB when one extends the calculation to higher order. The second order results of \cite{CMBdist,Clarkson1,Clarkson2}, developing from the detailed first order derivations in \cite{Durrer}, suggest that second order terms can become significant, of the order of 1\% or so. The reason that the distance to the CMB does not enter the angular power spectrum is not that its perturbation is very small, but that the CMB power spectrum is obtained by integrating brightness fluctuations over angles which leads to much smaller effects due to the  conservation of brightness under gravitational lensing~\cite{Clarkson1}.\\ 
 
 As regards supernove observations, we observe the luminosity distance $D_L(z) =(1+z)^2D_A$, where we can loosely use `distance' for both the angular diameter distance $D_A(z)$ or $D_L(z)$. Here $z$ is of course the true, perturbed redshift.
  The 1st and 2nd order perturbations to this expressions have been calculated in several papers: \cite{Bonvin:2005ps} (1st order), \cite{BenDayan:2012wi,Umeh:2014ana,CMBdist} 
   (2nd order).  Already at first order it was found that the variance of the distance from lensing is of the order of $10^{-3}-10^{-2}$, hence much larger than the Bardeen potential $\Psi \sim 10^{-5}$. The reason for this is that the dominant terms contain two derivatives of the Bardeen potential, $\kappa =  -\Delta\psi/2$, where $\psi$ is the lensing potential ${\cal O}(\Psi)$, and $\Delta$ is the Laplacian on the sphere.  Also a (formal) non-perturbative result has been derived~\cite{Fanizza:2013doa}. As can be seen in Fig.~1, the effect can add up to about 1\%, which contradicts the claims in KP. 
 \begin{figure}[h!]
 	\includegraphics[width=0.5\columnwidth]{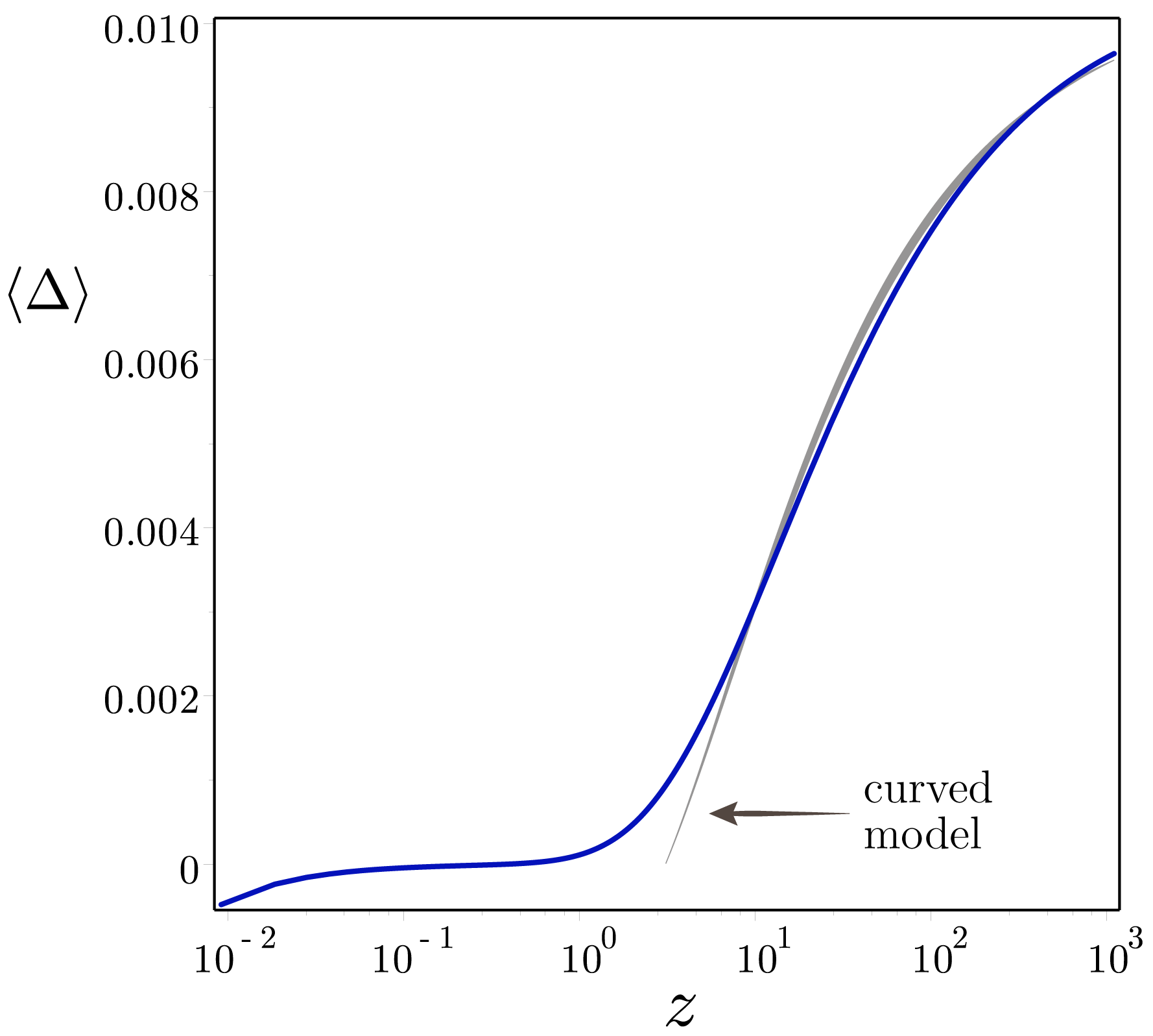} \includegraphics[width=0.5\columnwidth]{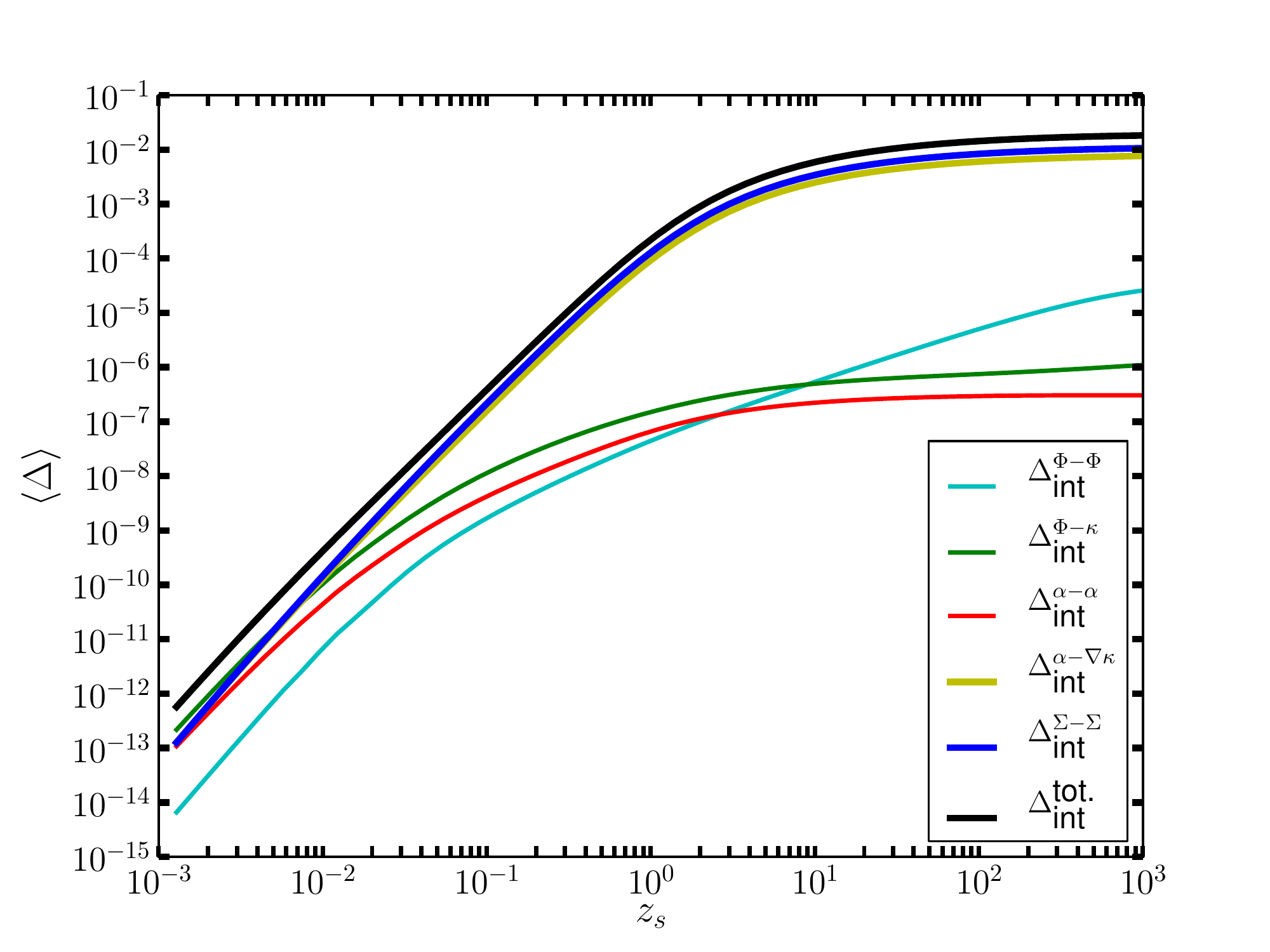}
 	\caption{Left:\textit{ Fractional correction $\langle\Delta\rangle(z)$ to the distance 
 			for a fiducial model $\Omega_m=0.3,h=0.68,h^2\Omega_b^2=0.0222,w=-1$ and $n_s=0.96$. The correction is negative for $z<0.25$, purely from the local contribution. At higher redshift the shift arises from the aggregated lensing term. 
 		For $z> 10$ the corrections grow  $\propto\chi_s^3$, and are similar to an open $\Lambda$CDM model with $\Omega_K^\text{eff}\approx 0.0066$ ({grey} `curved', shown for high $z$). Figure from }\cite{CMBdist}. 
 		Right: \textit{The different terms contributing to the distance correction.  
 		 Figure from }\cite{Umeh:2014ana}.}
 \end{figure}
 
 The `averaged' distance, averaged over directions, has also been discussed. It is important to notice that directional average and ensemble average do not commute at second order since also the surface element on the sphere is perturbed by the convergence $\kappa$   \cite{BenDayan:2013gc,Clarkson1,Clarkson2,Fleury:2016fda}. 
 
 What might be even more interesting is that the presence of structure in the Universe leads to irreducible statistical fluctuations, that is,  dispersion of these quantities. This has also been studied for the Hubble parameter, $H(z)$, see e.g.~\cite{Ben-Dayan:2014swa} and dispersions of order of 1-2\% or so have been found which have to be added to the observational error in the Hubble constant for local measurements.\\
 
  These results do not appear to accord with the results in KP. This could be due to the issues raised above.\\

 \textbf{Acknowledgments} We thank Nick Kaiser for a stimulating presentation of arguments in this regard at the \textit{CosmoBack} meeting held at LAM in Marseilles earlier this year, and Roy Maartens, Chris Clarkson, and Julien Larena for helpful comments on earlier versions of this note.



\begin{thebibliography}{99}
\bibitem{Bertotti}
Bertotti, B.  (1966) \href{https://www.jstor.org/stable/2415465?seq=1#page_scan_tab_contents}{``The luminosity of distant galaxies''} \textit{Proc. R. Soc. Lond}. \textbf{A 294}: 195-207.


\bibitem{Wei76}
Weinberg, S. (1976) \href{http://adsabs.harvard.edu/full/1976ApJ...208L...1W}{"Apparent luminosities in a locally inhomogeneous universe."} \textit{The Astrophysical Journal }\textbf{208}: L1-L3.


\bibitem{CMBdist}
Clarkson, C., Umeh, O., Maartens, R., and Durrer, R. (2014). ``What is the distance to the CMB?''. \textit{Journal of Cosmology and Astroparticle Physics} \textbf{11}: 036 \href{https://arxiv.org/abs/1405.7860}{https://arxiv.org/abs/1405.7860}


\bibitem{Kaiser_Peacock}
Kaiser, N, and Peacock, J A  (2015). ``On the bias of the distance-redshift relation from gravitational lensing.'' \textit{Monthly Notices of the Royal Astronomical Society} \textbf{455}: 4518-4547 \href{https://arxiv.org/abs/1503.08506}{https://arxiv.org/abs/1503.08506}

\bibitem{BE}
Ellis, G. F. R, and Bruni, M. (1989) \href{https://journals.aps.org/prd/abstract/10.1103/PhysRevD.40.1804}{``Covariant and gauge-invariant approach to cosmological density fluctuations}''. \textit{Physical Review} \textbf{D 40}, 1804.

\bibitem{Ehlers}
Ehlers, J (1961) Mainz paper: reprinted as \href{https://link.springer.com/article/10.1007/BF00759031}{``Contributions to the relativistic mechanics of continuous media.''} \textit{General Relativity and Gravitation} \textbf{25} (1993): 1225-1266.

\bibitem{ell71}
Ellis G F R (1971)  ``Relativistic cosmology''
in {\em General Relativity and Cosmology, Proceedings
	of the XLVII Enrico Fermi Summer School}, Ed. R K Sachs (Academic Press, New York), 104--182. See Ellis, George FR. \href{https://link.springer.com/article/10.1007/s10714-009-0760-7}{``Republication of: Relativistic cosmology.''} \textit{General Relativity and Gravitation}\textbf{ 41} (2009): 581-660.

\bibitem{Etherignton}
Ellis, G F R. (2007) \href{https://link.springer.com/article/10.1007%2Fs10714-006-0355-5}{``On the definition of distance in general relativity: IMH Etherington} (Philosophical Magazine ser. 7, vol. 15, 761 (1933)).'' \textit{General Relativity and Gravitation} \textbf{39}: 1047-1052.

\bibitem{Fanizza:2013doa}
Fanizza, G, Gasperini, M, Marozzi, G.~ and Veneziano,  G (2013),
``An exact Jacobi map in the  geodesic light-cone gauge  ''
\textit{JCAP} {\bf 1311}:019
\href{https://arxiv.org/abs/1308.4935}{https://arxiv.org/abs/1308.4935}

\bibitem{Sachs and Wolfe}
Sachs, R.K., Wolfe, A.M., Ellis, G., Ehlers, J. and Krasiński, A., (2007). \href{https://link.springer.com/article/10.1007%2Fs10714-007-0448-9}{``Republication of: Perturbations of a cosmological model and angular variations of the microwave background''} by RK Sachs and AM Wolfe. \textit{General Relativity and Gravitation} \textbf{39}:1929-1961.
	
\bibitem{Durrer}
Durrer R. (2008) \textit{The cosmic microwave background}. Cambridge: Cambridge University Press.

\bibitem{Clarkson1}
Bonvin, C , Clarkson, C, Durrer,  R, Maartens, and Umeh, O  (2015) 
``Do we care about the distance to the CMB? Clarifying the impact of second-order lensing'' 
\textit{JCAP} \textbf{2015}: 050 \href{https://arxiv.org/abs/1503.07831}{https://arxiv.org/abs/1503.07831}

\bibitem{Sachs:1961} 
Sachs, R.~K.~(1961)
\href{http://rspa.royalsocietypublishing.org/content/264/1318/309}{``Gravitational waves in general relativity. 6. The outgoing radiation condition,''}
Proc.\ Roy.\ Soc.\ Lond.\ A {\bf 264}, 309.

\bibitem{Pirani} Pirani, F A E (1965) In  \textit{Lectures on general relativity, Volume 1}, Trautmann, A, Bondi, H, and Pirani, F.A.E, (Prentice Hall).


\bibitem{Clarkson2}
Bonvin, C., Clarkson, C., Durrer, R., Maartens, R. and Umeh, O. (2015), ``Cosmological ensemble and directional averages of observables'', \textit{JCAP} \textbf{7}:40. \href{https://arxiv.org/abs/1504.01676}{https://arxiv.org/abs/1504.01676}


\bibitem{Lewis Challinor}
Lewis, A, and Challinor, A   (2006) "Weak gravitational lensing of the CMB." \textit{ Physics Reports} \textbf{429}: 1-65. \href{https://arxiv.org/abs/astro-ph/0601594}{https://arxiv.org/abs/astro-ph/0601594}

\bibitem{Bonvin:2005ps}
Bonvin, C, Durrer, R, and Gasparini, M.~A.~  (2006) 
``Fluctuations of the luminosity distance,'' 
\textit{Phys.\ Rev.}\ D {\bf 73}:023523
Erratum: [\textit{Phys.\ Rev.}\ D {\bf 85} (2012) 029901]
\href{https://arxiv.org/abs/astro-ph/0511183}{https://arxiv.org/abs/astro-ph/0511183}.

\bibitem{BenDayan:2012wi} Ben-Dayan, I.~, Marozzi, G.~, Nugier, F.~, and Veneziano, G.~  (2012),
``The second-order luminosity-redshift relation in a generic inhomogeneous cosmology,''
\textit{JCAP} {\bf 1211}:045
\href{https://arxiv.org/abs/1209.4326}{https://arxiv.org/abs/1209.4326}

\bibitem{Umeh:2014ana}
Umeh, O.~ Clarkson, C.~ and Maartens, R.~  (2014),
``Nonlinear relativistic corrections to cosmological distances, redshift and gravitational lensing magnification. II - Derivation,''
\textit{Class.\ Quant.\ Grav}.\  {\bf 31}:205001
\href{https://arxiv.org/abs/1402.1933}{https://arxiv.org/abs/1402.1933}





\bibitem{BenDayan:2013gc}
Ben-Dayan, I.~, Gasperini, M.~, Marozzi, G.~, Nugier, F.~, and Veneziano, G.~  (2013),
``Average and dispersion of the luminosity-redshift relation in the  concordance model,''
\textit{JCAP} {\bf 1306}:002
\href{https://arxiv.org/abs/1308.4935}{https://arxiv.org/abs/1308.4935}
%


\bibitem{Fleury:2016fda}
Fleury, P.~ Clarkson C.~and Maartens R.~(2017),
``How does the cosmic large-scale structure bias the Hubble diagram? ,''
\textit{JCAP} {\bf 1703}:  062
\href{https://arxiv.org/abs/1612.03726}{https://arxiv.org/abs/1612.03726}

\bibitem{Ben-Dayan:2014swa}
Ben-Dayan, I.,~~Durrer, R., Marozzi G., ~and ~Schwarz D.~J.(2014),
``The value of $H_0$ in the inhomogeneous Universe,''
\textit{Phys.\ Rev.\ Lett.}\  {\bf 112}:  221301
\href{https://arxiv.org/abs/1401.7973}{https://arxiv.org/abs/1401.7973}




\end{thebibliography}
  \end{document}